# Anisotropic Interfacial Force Field for Interfaces of Water with Hexagonal Boron Nitride

Zhicheng Feng,[1,#] Zhangke Lei,[1,#] Yuanpeng Yao,[1] Jianxin Liu,[1] Bozhao Wu,[1,*] and Wengen Ouyang[1,2,*]

[1]*Department of Engineering Mechanics, School of Civil Engineering, Wuhan University, Wuhan, Hubei 430072, China.*

[2]*State Key Laboratory of Water Resources & Hydropower Engineering Science, Wuhan University, Wuhan, Hubei 430072, China.*

[#]These authors contributed equally to this work.

[*]Corresponding authors. Email: bozhwu@whu.edu.cn; w.g.ouyang@whu.edu.cn

ABSTRACT

This study introduces an anisotropic interfacial potential that provides an accurate description of the van der Waals (vdW) interactions between water and hexagonal boron nitride (*h*-BN) at their interface. Benchmarked against the strongly constrained and appropriately normed (SCAN) functional, the developed force field demonstrates remarkable consistency with reference data sets, including binding energy curves and sliding potential energy surfaces for various configurations involving a water molecule adsorbed atop the *h*-BN surface. These findings highlight the significant improvement achieved by the developed force field in empirically describing the anisotropic vdW interactions of the water/*h*-BN heterointerfaces. Utilizing this anisotropic force field, molecular dynamics simulations demonstrate that atomically-flat pristine *h*-BN exhibits inherent hydrophobicity. However, when atomic-step surface roughness is introduced, the wettability of *h*-BN undergoes a significant change, leading to a hydrophilic nature. The calculated water contact angle (WCA) for the roughened h-BN surface is approximately 64°, which closely aligns with experimental WCA values ranging from 52° to 67°. These findings indicate the high probability of the presence of atomic steps on the surfaces of experimental *h*-BN samples, emphasizing the need for further experimental verification. The development of the anisotropic interfacial force field for accurately describing interactions at the water/*h*-BN heterointerfaces is a significant advancement in accurately simulating the wettability of two-dimensional (2D) materials, offering a reliable tool for studying the dynamic and transport properties of water at these interfaces, with implications for materials science and nanotechnology.

# 1. Introduction

Hexagonal boron nitride (*h*-BN) has generated significant interest in various fields, including water cleaning,[1, 2] seawater desalination,[3-5] biomedicines[6, 7], nanopore DNA sequencing,[8, 9] fuel cells,[10, 11] and osmotic power harvesting,[12, 13] among others. In these applications, there are intimate and non-negligible interactions between water and *h*-BN nanopores, nanochannels and films, necessitating a comprehensive understanding of *h*-BN's wettability. However, the intrinsic wettability of *h*-BN has remained elusive over the past few decades, as evidenced by the wide range ($40° - 165°$) of reported water contact angles (WCAs) for monolayer and bulk *h*-BN in experiments.[1, 14-18] Several factors contribute to this complexity, including surface contamination, surface roughness, and the presence of defects within *h*-BN layers.[15, 17, 19] Recent experimental endeavors have focused on reducing the influence of airborne contamination by measuring WCAs of high-quality *h*-BN films under controlled conditions, revealing a hydrophilic nature with WCAs falling within the range of $52° - 67°$.[15, 18, 20, 21] It is worth noting that experimental high-quality *h*-BN films may not be perfectly atomically flat and can contain surface defects or steps.[15, 18] These imperfections have the potential to significantly influence the wettability of *h*-BN.[14, 19, 22]

In addition to experimental investigations, molecular dynamics (MD) simulations have emerged as a valuable tool for gaining insights into the wettability of two-dimensional (2D) materials such as graphene and *h*-BN.[23-42] Through MD simulations, researchers have been able to explore and understand the intricate dynamics and behavior of water molecules on these surfaces. However, the accuracy of these simulations relies heavily on the chosen force fields. In the case of describing the van der Waals (vdW) interaction between water molecules and *h*-BN, the commonly used isotropic Lennard-Jones (LJ) potential ($V(r) = 4\epsilon[(\sigma/r)^{12} - (\sigma/r)^6]$) has been widely adopted.[19, 23-27, 43-45] These simulations predict a wide range of WCA for *h*-BN ($0° - 86°$).[23-27, 44-46] The typical parameters describing the interactions between oxygen (O) atoms in water ($H_2O$) and boron (B) / nitride (N) atoms in *h*-BN, e.g., $\varepsilon_{BO}$ and $\varepsilon_{NO}$, fall within the range of $2.16 - 6.13$ meV and $4.33 - 8.97$ meV, respectively.[24, 25, 44, 45] The wide range of LJ parameters leads to inconsistent WCAs, indicating that the LJ potential is not suitable for studying the wetting properties of *h*-BN.

Density functional theory (DFT) calculations provide a more accurate description of $H_2O$/*h*-BN interactions. However, it is crucial to carefully select a DFT method with the appropriate vdW correction to obtain reliable results.[47-49] Recent researches have demonstrated that several many-body calculation methods, including coupled cluster theory (CCSD(T)), diffusion Monte Carlo (DMC) method, lattice regularized DMC (LRDMC), second order Møller-Plesset (MP2), random phase approximation (RPA), and strongly constrained and appropriately normed (SCAN) functional, can predict the interaction strength of $H_2O$/$h$-BN with sub-chemical accuracy.[50, 51] Nevertheless, using DFT methods directly for calculating the WCA of $h$-BN systems is not feasible due to their computationally intensive nature. Consequently, accurately and efficiently describing the intrinsic wetting properties of large-area $h$-BN films remain a challenging task. To overcome this limitation, one possible approach is to develop an empirical force field capable of accurately describing the interaction between water molecules and $h$-BN within the framework of MD simulations. Inspired by the successful application of the registry-dependent interlayer potential to two-dimensional (2D) layered materials,[52-58] we developed anisotropic interfacial potential (AIP) to accurately describe the interaction between water molecules and $h$-BN in the current work. In our previous work investigating the water/graphene heterointerface,[59] we demonstrated that the accuracy of the developed AIP strongly depends on the accuracy of the DFT reference data used for parameterization. In a recent study by Al-Hamdani et al., the performance of various DFT exchange-correlation functionals, including CCSD(T), LRDMC, DMC, MP2, and RPA methods,[50, 51] was investigated in terms of their ability to accurately predict the binding energy (BE) of water molecules on the $h$-BN surface. The results revealed that most of the considered functionals, which incorporate van der Waals (vdW) interactions, significantly overestimated the BE of water molecules on $h$-BN, except for the SCAN functional within a meta-generalized gradient approximation (meta-GGA). The SCAN functional, despite lacking explicit vdW correction, exhibited remarkable accuracy in predicting the BE, closely aligning with the DMC results of $H_2O$/boronene and $H_2O$/$h$-BN (as shown in Table 1).[50, 51] Therefore, we utilized the SCAN semi-local density functional to conduct DFT reference calculations for $H_2O$/$h$-BN heterostructures. Here, BE curves and sliding potential energy surfaces (PESs) for various $H_2O$/$h$-BN configurations are calculated to capture the anisotropic nature of the vdW interaction between the water molecule and $h$-BN. The importance of sliding PESs for accurately investigating the vdW interaction between water and 2D materials, which was overlooked in most previous theoretical studies, has been pointed out in recent studies.[43, 59] The inclusion of sliding PESs in force field development is crucial for accurately investigating the transport and tribological characteristics of water on 2D materials.[59]

In this study, we demonstrate that the AIP achieves excellent agreement with all the DFT reference data sets, highlighting its capability to accurately describe the vdW interaction between water and $h$-BN heterointerfaces. In contrast, the LJ potential fails to capture the anisotropic characteristics of the vdW interaction in these systems, as it cannot adequately parameterize two BE curves with a single set of parameters, let alone capturing the behavior of sliding PESs.[24, 59] These results highlight the superiority of our AIP in providing a more self-consistent and reliable description of the vdW interaction between water molecules and *h*-BN compared to the LJ potential. When utilizing the AIP, we obtained a calculated WCA of approximately 115° for atomically-flat pristine *h*-BN, indicating its hydrophobic nature. However, this prediction contradicts experimental observations. Kumar Verma and Seal et al.[19, 43] claimed that this inconsistency may be attributed to the presence of defects or surface roughness/steps in experimental *h*-BN samples. Their further investigations confirm that the surface edges of *h*-BN significantly reduce the calculated WCA, leading to a hydrophilic nature that aligns with the experimental results. Inspired by their work, we performed MD simulations of water on the *h*-BN surface with atomic steps and obtained a WCA of 64°, which agrees well with the experimental results.[15, 18, 20, 21] These findings support the notion that surface roughness plays a crucial role in the wetting behavior of water on the *h*-BN surface.

## 2. Methodology

### 2.1. DFT Method

The DFT calculations were performed using the Vienna Ab initio Simulation Package (VASP) with the projector augmented wave (PAW) method for the ion-electron interaction,[60] using a kinetic energy of 500 eV. The exchange-correlation functional was treated using the SCAN functional with a meta-generalized gradient approximation (meta-GGA)[61]. To mitigate image interactions between water molecules in the basal plane, a supercell of the $H_2O$/*h*-BN models was set as $4a \times 4a \times 1$ (where $a = 2.489$ Å is the lattice constant of *h*-BN in the primitive cell). Likewise, to avoid image interactions along the *z*-direction, a vacuum thickness of 30 Å was utilized for all configurations. A $5 \times 5 \times 1$, Γ-centered mesh grid was used for Brillouin-zone integration. Convergence tests were conducted on the DFT results to assess the influence of the supercell size, vacuum size, and *k*-points (see Section S1 of the Supporting Information (SI) for further details).

### 2.2. Model Systems

For DFT calculations, the $H_2O$/*h*-BN heterogeneous configurations consisted of a water molecule adsorbed atop monolayer *h*-BN. To begin with, we chose six distinct $H_2O$/*h*-BN configurations, which encompassed a variety of stacking modes and orientations for a water molecule on top of monolayer *h*-BN, as detailed in Section S2 of the SI. These configurations were subsequently optimized using the SCAN functional, resulting in two meta-stable configurations: *i*) the "one-leg" configuration (Figs. 1(a-b)), involves the water molecule's O atom positioned approximately on top of a N atom of *h*-BN, with one H atom pointing downwards towards the same N atom; *ii*) the "zero-leg" configuration (Figs. 1(c-d)), has the O atom atop a B atom, and the plane of the water molecule approximately parallel to the *h*-BN plane. DFT calculations demonstrate that the "one-leg" configuration is the most stable configuration with the lowest BE between water and *h*-BN, which is consistent with previous reports.[25, 50, 51] The bond lengths of B-N and O-H were determined to be 1.436 Å and 0.964 Å, respectively, and the H-O-H bond angle was calculated to be 105.16°.

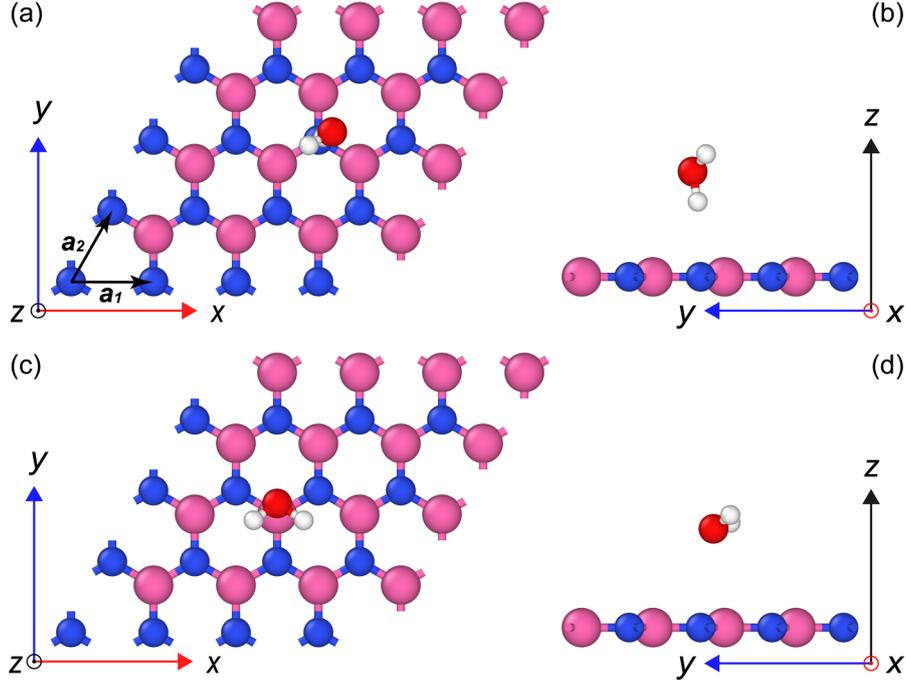

**FIG. 1.** *The two meta-stable water configurations adsorbed on h-BN surface, obtained from the optimizations at the SCAN functional level. The top (a, c) and side (b, d) views of the "one-leg" (top row) and "zero-leg" (bottom row) configurations are provided. The magnitude of the lattice constants $a_1$, $a_2$ in the unit cell of h-BN is 2.487 Å. Magenta, blue, red and white spheres represent N, B, O and H atoms, respectively.*

## 2.3. DFT reference data

### 2.3.1. BE Curves Calculations

Using the two meta-stable configurations depicted in Fig. 1, we obtained two BE curves by varying the distance between the water molecule's O atom and the *h*-BN surface from 2 Å to 16 Å. The calculated BEs (and the corresponding equilibrium distances) for the "one-leg" and "zero-leg" configurations are 99.3 meV (3.2 Å) and 67.2 meV (3.0 Å), respectively. Our calculated BEs using the SCAN functional are close to the values obtained using the DMC method, which is consistent with the findings of Al-Hamdani et al.[51] Table 1 provides a detailed comparison of the calculated BEs using different functionals.

**TABLE 1.** Binding energy ($E_b$) and equilibrium distances ($d_{eq}$, defined as the O/$h$-BN vertical distance) for the "one-leg" and "zero-leg" $H_2O$ model (see Fig. 1) atop monolayer $h$-BN, obtained through different computational methods.

| Model name | "One leg" | | "Zero leg" | |
| --- | --- | --- | --- | --- |
| Approach | $E_b$ (meV) | $d_{eq}$ (Å) | $E_b$ (meV) | $d_{eq}$ (Å) |
| LRDMC[51] | 107±7 | 3.40 | -- | -- |
| MP2[51] | 110 | 3.25 | -- | -- |
| DMC[51] | 95±5 | 3.40 | -- | -- |
| RPA[51] | 89 | 3.36 | -- | -- |
| PBE[50] | 44 | 3.40 | 23 | 3.40 |
| PBE+D3[50] | 143 | 3.20 | 105 | 3.00 |
| PBE+MBD[51] | 146 | 3.25 | -- | -- |
| optB86b-vdW[50] | 168 | 3.20 | 142 | 3.00 |
| SCAN (This work) | 99.3 | 3.18 | 67.2 | 3.04 |

To account for the possible occurrence of various orientations of water molecules at the $H_2O$/$h$-BN heterointerface in the presence of hydrogen bonding among water molecules, additional stacking configurations were considered. Specifically, the "one-leg", "two-leg" (see the upper inset of Fig. 2(c)), and "zero-leg" configurations were rotated along a single rotational axis to produce multiple distinct configurations, from which three BE curves related to the rotation angle were obtained (Figs. 2(b-c)). It's worth noting that rotational symmetries exist for the water molecule in the "two-leg" and "zero-leg" configurations, with sixfold and threefold symmetries, respectively (see the insets of Fig. 2(c)), while the "one-leg" configuration has no rotational symmetry (see the inset of Fig. 2(b)). Therefore, the rotational angles of the water molecule around the axis in the "one-leg", "two-leg" and "zero-leg" configurations are 360° (Fig. 2(b)), 60°, and 120° (Fig. 2(c)), respectively. Throughout the rotational process, the vertical distance between the O atom and the $h$-BN sheet remains fixed at 3.2 Å for the "one-leg" and "two-leg" $H_2O$ models, and 3.0 Å for the "zero-leg" configuration. The results in Figs. 2(b-c) indicate that the rotating BE of the $H_2O$/$h$-BN heterostructure is highly dependent on the orientation of the $H_2O$ molecule. This finding shows the incapability of the LJ potential to precisely describe the vdW interaction between water and $h$-BN, as explained in Section S3 of the SI. On the contrary, the AIP fitting exhibits remarkable consistency with the calculated data sets using the SCAN functional, as evidenced in Fig. 2.

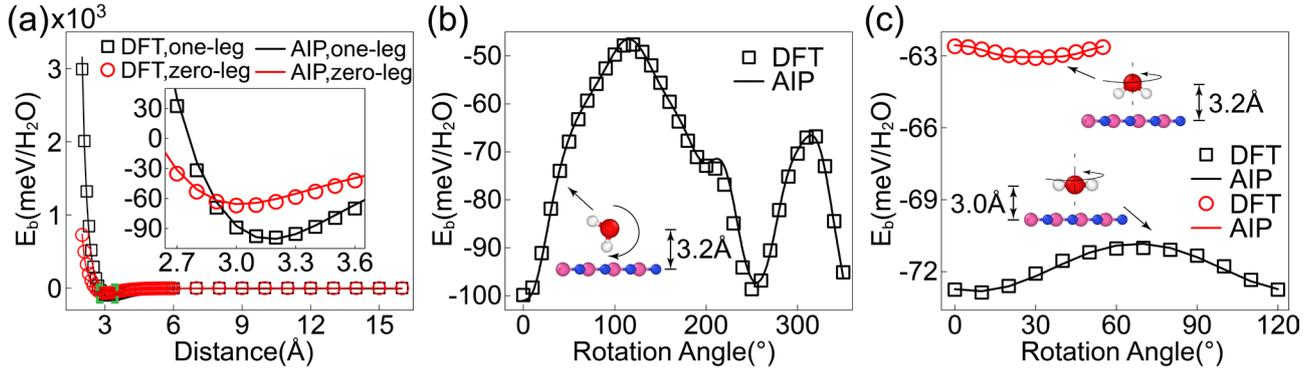

**FIG. 2.** *BE curves of a water molecule adsorbed on h-BN surface. (a) BE as a function of the interlayer distance for "one-leg" (marked in black) and "zero-leg" (marked in red) $H_2O$ models, calculated by using SCAN method (open symbols) and AIP (solid lines), respectively. The inset offers zoom-in on the equilibrium interlayer distance region. (b, c) Rotating BEs calculated through rotations of the (b) "one-leg", (c) "zero-leg" (marked in black) and "two-leg" (marked in red) $H_2O$ models. The insets of (b,c) illustrate the directions of rotations as well as O/h-BN vertical distances.*

*2.3.2. Sliding PESs calculations*

We computed the sliding PESs by rigidly moving the water molecule parallel to the *h*-BN layer along the armchair and zigzag directions of the *h*-BN for both the "one-leg" and "zero-leg" configurations at their equilibrium distances. As shown in Fig. 3, the patterns of sliding PESs obtained with DFT and AIP are highly similar. Specifically, the maximal energy corrugations of the DFT (AIP) PESs for the "one-leg" and "zero-leg" $H_2O$ configurations are 55.9 meV/$H_2O$ (54.8 meV/$H_2O$) and 42.1 meV/atom (43.2 meV/atom), respectively. The maximum BE differences between AIP and DFT results among all these configurations are 1.9 meV/$H_2O$ (3.5%) and 2.8 meV/$H_2O$ (6.7%) for the "one-leg" and "zero-leg" $H_2O$/*h*-BN models, respectively, as illustrated in Figs. 3(c,f). These results indicate that the sliding PES calculated using our developed AIP (Figs. 3(b,e)) is in good agreement (both qualitatively and quantitatively) with the DFT reference data obtained using the SCAN functional (Figs. 3(a,d)). Notably, we checked that the LJ potential can describe only one of the BE curves shown in Fig. 2(a), and the magnitude of the sliding PES corrugation predicted by LJ is severely underestimated (see Section S3 in the SI for details). The aforementioned results clearly demonstrate the ability of the AIP to accurately describe the anisotropic interaction of $H_2O$/*h*-BN heterointerfaces.

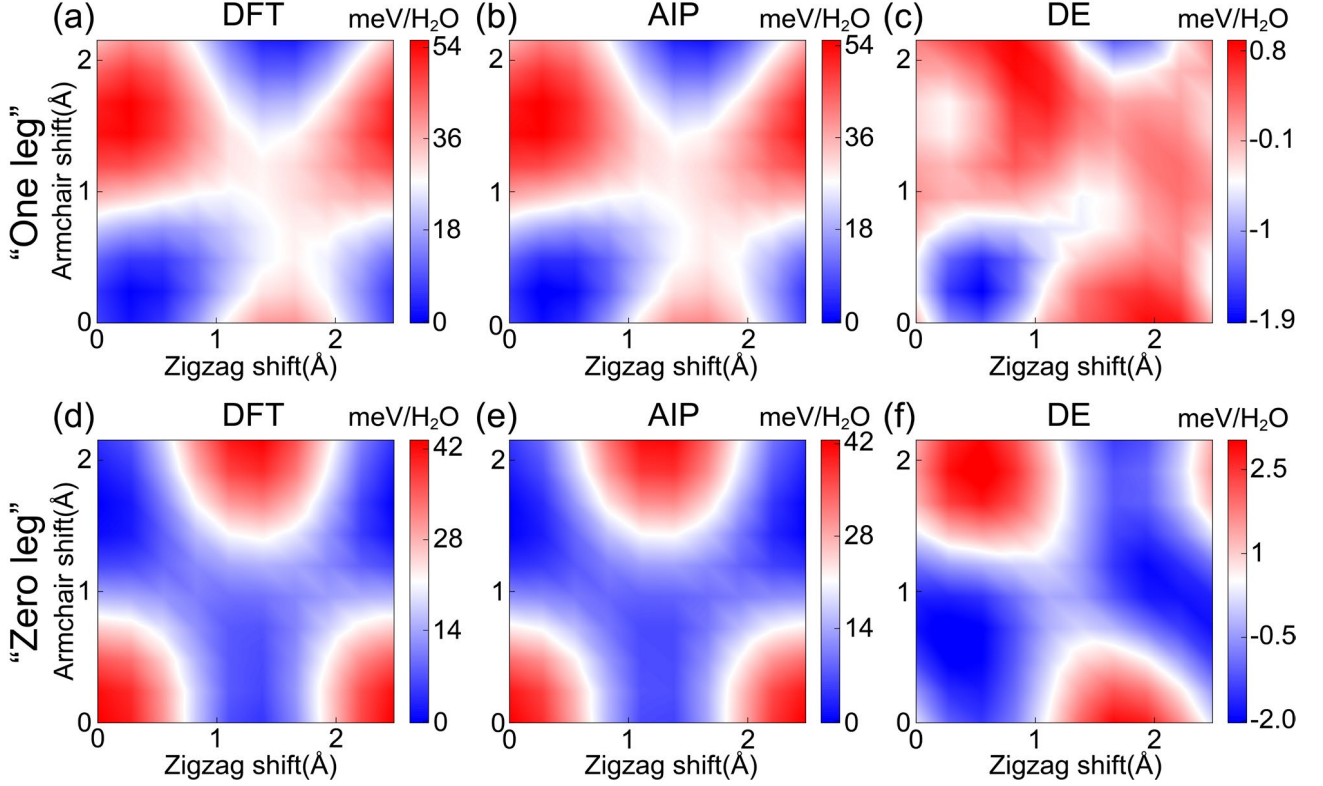

*FIG. 3. Sliding PESs of "one-leg" (upper rows) and "zero-leg" (lower rows) water configurations on h-BN surface at interlayer distances of 3.2 Å and 3.0 Å, respectively. The PESs were predicted using the SCAN functional at DFT level (a, d) and the parameterized AIP (b, e), respectively. The differences of the DFT reference data and AIP results are given in panels (c, f).*

## 3. Structure of the Anisotropic Interfacial Potential

To accurately describe the $H_2O$/$h$-BN interaction, it is necessary for the developed AIP to effectively capture both the BE curves and sliding PESs, while also accurately representing the anisotropic nature of the interactions between water and $h$-BN. We developed the AIP based on our previous experience in developing anisotropic interlayer potentials for 2D and water/2D materials, all of which demonstrated significant improvements over conventional isotropic force fields.[58, 59, 62-65] The AIP describing the $H_2O$/$h$-BN interaction consists of three components: the anisotropic short-range Pauli repulsive term $E_{\text{Rep}}(\mathbf{r}_{ij}, \mathbf{n}_i, \mathbf{n}_j)$, the isotropic long-range van der Waals attractive term $E_{\text{Att}}(r_{ij})$, and the electrostatic interaction $E_{\text{Coul}}(r_{ij})$:

$$E(\mathbf{r}_{ij}, \mathbf{n}_i, \mathbf{n}_j) = \text{Tap}(r_{ij})[E_{\text{Rep}}(\mathbf{r}_{ij}, \mathbf{n}_i, \mathbf{n}_j) + E_{\text{Att}}(r_{ij}) + E_{\text{Coul}}(r_{ij})] \qquad (1)$$

The taper cutoff function $\text{Tap}(r_{ij})$ incorporated in the formula provides a smooth and continuous long-range cutoff at a distance of $r_{ij} = R_{\text{cut}}$, allowing for up to the third-order derivative of the potential. In our experience of developing water/2D material force fields,[59] incorporating the Tap function greatly alleviates the computational stress while simultaneously offering a satisfactory description for the interface interactions.[66]

$$\text{Tap}(r_{ij}) = \frac{20}{R_{\text{cut}}^7}r_{ij}^7 - \frac{70}{R_{\text{cut}}^6}r_{ij}^6 + \frac{84}{R_{\text{cut}}^5}r_{ij}^5 - \frac{35}{R_{\text{cut}}^4}r_{ij}^4 + 1 \qquad (2)$$

*3.1. Short-Range Repulsion*

Recent researches have documented the noteworthy impact of electron cloud overlap between water molecules and 2D materials, such as graphene and *h*-BN, on their wettability.[47, 67] However, the commonly employed isotropic LJ potential fails to account for the effect of electron cloud overlap, resulting in an inaccurate description of the BE near the equilibrium interlayer distance,[65, 68] as reported in a recent research on $H_2O$/*h*-BN systems.[24] In order to precisely describe the anisotropic Pauli repulsions that arise from the electron cloud overlap at the $H_2O$/*h*-BN heterointerface, we utilized the scheme developed by Kolmogorov and Crespi,[65] which contains a Morse-like exponential isotropic term multiplied by an anisotropic correction with the following form:

$$E_{\text{Rep}}(\mathbf{r}_{ij}, \mathbf{n}_i, \mathbf{n}_j) = \text{Tap}(r_{ij})e^{\alpha_{ij}\left(1-\frac{r_{ij}}{\beta_{ij}}\right)}\left[\epsilon_{ij} + C_{ij}\left(e^{-\left(\frac{\rho_{ij}}{\gamma_{ij}}\right)^2} + e^{-\left(\frac{\rho_{ji}}{\gamma_{ij}}\right)^2}\right)\right] \qquad (3)$$

Here, $\varepsilon_{ij}$ and $C_{ij}$ are constant factors that establish the energy scales related to the isotropic and anisotropic repulsion, respectively, $\beta_{ij}$ and $\alpha_{ij}$ determine the interaction ranges associated with these factors, while $\alpha_{ij}$ is a parameter that governs the steepness of the isotropic repulsion function. The repulsive term requires information regarding both the actual ($r_{ij}$) and the lateral ($\rho_{ij}$) distances between the $i$ and $j$ atomic sites. The normalized normal vector $\mathbf{n}_i$ (i.e., $||\mathbf{n}_i||= 1$) is utilized to determine the lateral distance $\rho_{ij}$ between the B (N) atom $i$ in the *h*-BN layer and the H (O) atom $j$ from water molecules:

$$\rho_{ij}^2 = r_{ij}^2 - (\mathbf{n}_i \cdot \mathbf{r}_{ij})^2$$
$$\rho_{ji}^2 = r_{ij}^2 - (\mathbf{n}_j \cdot \mathbf{r}_{ij})^2 \qquad (4)$$

In which $\mathbf{n}_i$ defines the local normal direction to the *h*-BN sheet (or to the water molecule) at the position of atom $i$ (see Fig. 4(a)). The normal vector of a B (N) atom is determined as a vector perpendicular to the triangle created by its three closest neighbors.[55, 65] Fig. 4(c) illustrates the differential charge density that denotes the electron transfer between a water molecule and *h*-BN layer, and this charge redistribution near the equilibrium position leads to the strong anisotropic characteristic between water molecules and *h*-BN, e.g., the strong dependence of BE on the orientation of the water molecule. To accommodate the anisotropic characteristics at the interface of the water molecule and *h*-BN, we assumed that the H atoms' normal vectors align with their corresponding O-H bonds, while the O atom's normal vector is defined as the average of both H atoms' normal vectors (see Fig. 4(b)):[59]

$$\begin{cases} \boldsymbol{n}_{H_j} = \boldsymbol{r}_{\overrightarrow{OH_j}}/\left|\boldsymbol{r}_{\overrightarrow{OH_j}}\right|, & j = 1,2 \\ \boldsymbol{n}_O = (\boldsymbol{r}_{\overrightarrow{OH_1}} + \boldsymbol{r}_{\overrightarrow{OH_2}})/\left|\boldsymbol{r}_{\overrightarrow{OH_1}} + \boldsymbol{r}_{\overrightarrow{OH_2}}\right| \end{cases} \quad (5)$$

*3.2. Long-Range Attraction*

The long-range vdW attraction term incorporated in the AIP is inspired by the dispersion correction scheme developed by Tkatchenko and Scheffler.[69] This methodology enhances standard exchange-correlation density functional approximations, which are inadequate in capturing long-range vdW interactions, by introducing LJ-type pairwise terms that diminish as $C_6/r^6$ with the interatomic distance, denoted as $r_{ij}$. These terms are appropriately damped in the short-range to prevent double-counting of short-range correlation effects. The pairwise long-range attraction terms can be expressed in the following form:

$$E_{\text{Att}}(r_{ij}) = \text{Tap}\,(r_{ij})\left\{-\frac{1}{1+e^{-d\left[\left(r_{ij}/(s_R \cdot r_{ij}^{\text{eff}})\right)-1\right]}} \cdot \frac{C_{6,ij}}{r_{ij}^6}\right\} \quad (6)$$

where $r_{ij}$ is the distance between atom $i$ (either B or N) and atom $j$ (either H or an O). The parameters $d$ and $s_R$, both unitless, determine the slope and initiation point of the short-range Fermi-type dampening function. Additionally, $r_{ij}^{\text{eff}}$ denotes cumulative effective radii of atom $i$ and $j$, while $C_{6,ij}$ corresponds to the pairwise dispersion coefficients.[69]

*3.3. Electrostatic Interaction*

In contrast to graphene, *h*-BN is a polarized 2D material. Since the water molecules are also polarized, the Coulomb interaction between water molecules and *h*-BN cannot be neglected. To incorporate this interaction, which arises from their electronic polarization, we employ the following expression:[64]

$$E_{\text{Coul}}(r_{ij}) = \text{Tap}(r_{ij}) \left[ kq_iq_j / \sqrt[3]{r_{ij}^3 + (1/\lambda_{ij})^3} \right] \tag{7}$$

Here, $k = 14.399645 \text{eV} \cdot \text{Å} \cdot \text{C}^{-2}$ is Coulomb's constant, and $q_i$ and $q_j$ are the effective charges of B (N) atom $i$ from $h$-BN and H (O) atom $j$ from water molecules, given in units of the absolute value of the electron charge. $\lambda_{ij}$ stands as a shielding parameter utilized to mitigate the presence of a short-range singularity in the electrostatic interaction in regions where the interlayer potential is dominated by Pauli repulsions between overlapping electron clouds. In this study, we adopted the fixed effective atomic charge approximation for pristine $h$-BN, with $q_B = 0.42e$ and $q_N = -0.42e$.[62] We adopted the atomic charges from the TIP4P/2005 water model,[70] where the O atom has a charge of $q_O = -1.1128\,e$ and each H atom carries a charge of $q_H = 0.5564e$.[70]

To improve the overall accuracy of the fitting results, especially near the equilibrium position, we used an interior-point algorithm and introduced weighting factors for different DFT reference data in the fitting procedure, which is carried out using MATLAB.[71,72] Details of the fitting procedure and the fitted AIP parameters are provided in Section S4 of the SI.

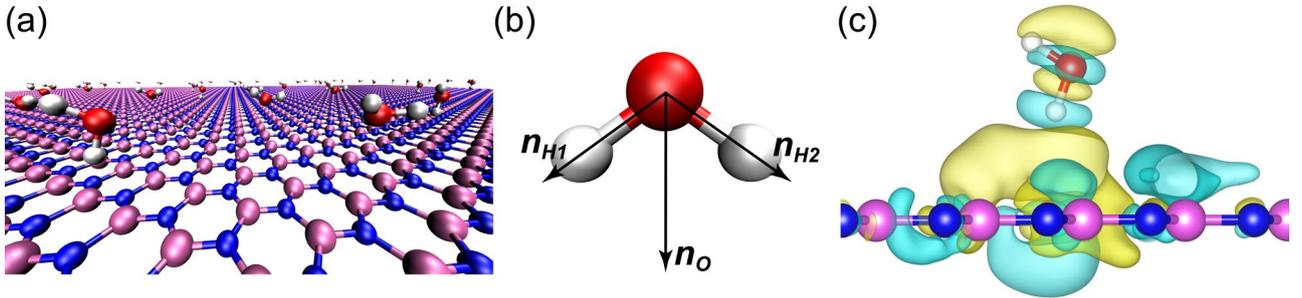

*FIG. 4. (a) Illustration of water molecules adsorbed on h-BN surface, (b) definition of local normal vectors $\mathbf{n}_O$ and $\mathbf{n}_H$ for water molecules, (c) visualization of electron density difference upon deposition of single "one-leg" water molecule on monolayer h-BN. The yellow and cyan clouds signify accumulation and depletion of electron, respectively.*

## 4. The Water Contact Angle of *h*-BN Systems

In this section, we conducted comprehensive MD simulations to investigate the wettability characteristics of both monolayer and multilayer $h$-BN. The main focus was to calculate the water contact angles (WCAs) on these surfaces. Before conducting the simulations, we integrated the developed AIP into the widely used LAMMPS code,[73] which allows us to accurately model the interactions between water molecules and *h*-BN at a large-scale. We considered two models: atomically-flat pristine *h*-BN (Fig. 5(a)) and a roughened *h*-BN surface (Fig. 5(c)) with atomic step edges.[19] These models allowed us to investigate the effects of surface roughness on the wetting characteristics of water molecules on *h*-BN. Our MD simulations were carried out under the NVT ensemble using the Nosé-Hoover thermostat[74] to maintain a constant temperature of 300 K. Snapshots of atomic configurations were visualized using the Open Visualization Tool (OVITO).[75] In all MD simulations, we employed the TIP4P/2005 water model,[70] which has been widely recognized for its ability to strike a balance between accuracy and computational efficiency in investigating the wettability of 2D materials.[31, 36, 59, 76-81] The SHAKE algorithm[82] was utilized to maintain the structural rigidity of water molecules. The cutoff for LJ and electrostatic interactions among water molecules was set as 12 Å. The long-range electrostatic interactions among water molecules were computed using the particle-particle-particle-mesh (PPPM) method.[83] The interfacial interactions between water molecules and *h*-BN were described by the AIP with a cutoff of 16 Å. To optimize computational efficiency, we treated the atoms of h-BN as rigid bodies during the simulations, without integrating their motion. Previous studies have shown that the flexibility of the graphene substrate has a negligible impact on the calculated WCA.[59] In all MD simulations, the velocity-Verlet integrator was employed with a time step of 2 fs, which has been proven to be sufficiently small to obtain convergent results.[84]

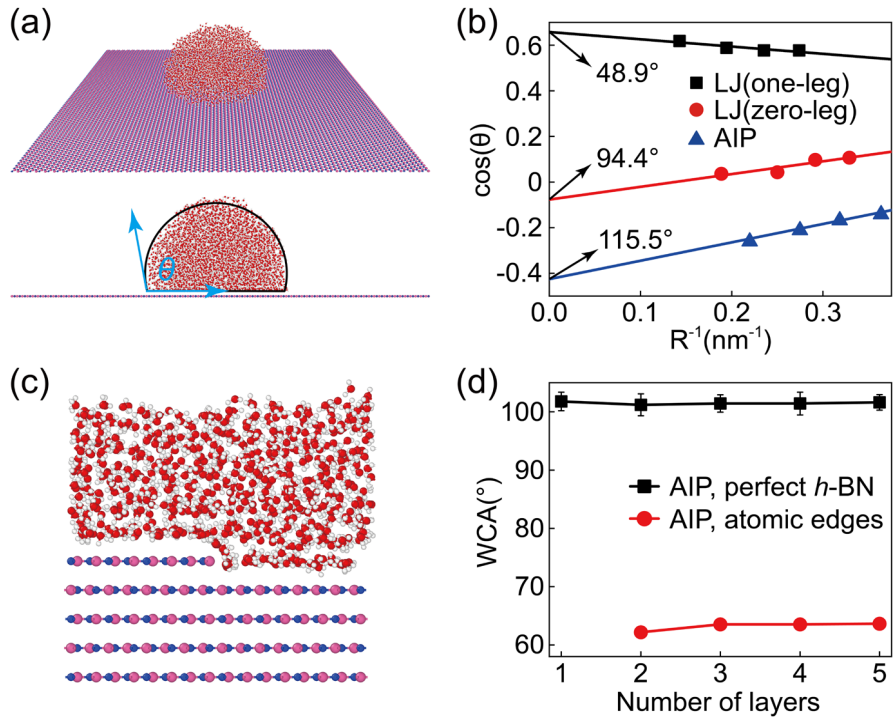

*FIG. 5.* Investigations on the wettability of h-BN based on our developed AIP. (a) Snapshot of a water droplet on atomically-flat pristine monolayer h-BN when reached equilibrium (top: perspective view; bottom: side view), which contains 4000 water molecules ($N_w = 4000$). (b) WCAs ($\theta$) for droplets

*containing different numbers of water molecules ($N_w$ = 1536, 2312, 4000, and 9464) on monolayer h-BN, from which the $\theta_\infty$ for an infinitely large droplet can be extrapolated using Eq. (8). (c) Adhesion work calculation setup for water molecules on bulk h-BN with atomic step edges using FEP method. (d) Effect of h-BN layers' thickness on the WCA.*

*4.1. Pristine Monolayer h-BN*

First, we calculated the WCA of a water droplet depositing on a monolayer pristine *h*-BN substrate using the radial density profile method (Fig. 5(a)). This approach has been previously validated and demonstrated to yield reliable results, particularly when applied in conjunction with the TIP4P/2005 water model as employed in our previous study.[59] In MD simulations, periodic boundary conditions were applied for the lateral directions of the *h*-BN substrate and a simulation box with dimensions of 19.816 nm × 201.455 nm × 15.000 nm was employed to ensure the exclusion of self-interactions between water molecule images. It is known that the WCA of *h*-BN depends on the size of the deposited droplet.[85] To determine the macroscopic contact angle $\theta_\infty$, model systems with 1536, 2312, 4000, and 9464 water molecules ($N_w$) on the surface of monolayer pristine *h*-BN were considered. To ensure stable droplet shapes and avoid errors arising from initial instabilities, data collection was initiated after a 5 ns equilibration period. According to the density profile method for calculating WCAs, we can obtain the contour of the droplet by interpolation, which yields the WCA after fitting a circle (further details can be found in Ref. 59). The obtained water contact angles (WCAs) were then used to fit the line-tension modified Young equation for systems with different droplet sizes:[31, 76, 78, 80]

$$\cos(\theta_R) = \cos(\theta_\infty) - \frac{1}{\gamma_{lv}} \frac{\tau}{R} \tag{8}$$

Here, $R$ represents the average radius of the contact area between the droplet and the *h*-BN sheet. $\gamma_{lv}$ and $\tau$ correspond to the surface tension and line tension of water, respectively. $\theta_R$ is the measured WCA for a given droplet, and $\theta_\infty$ is the contact angle of an infinite water droplet. By fitting Eq. (8), we obtained $\theta_\infty = 115.5°$ for monolayer pristine *h*-BN. As a comparison, we made an effort to parameterize the LJ potential based on the SCAN calculated data. However, the LJ potential can fit only one of the BE curves for the "one-leg" or "zero-leg" H₂O model, not to mention fitting the sliding PESs (see Section S3 of the SI). Consequently, the LJ potential was parameterized using the two BE curves (see Fig. 2(a)) separately, resulting in two distinct sets of LJ parameters (see Table S2 of the SI). Correspondingly, the calculated $\theta_\infty$ using the reparametrized LJ potential for the monolayer pristine *h*-BN was 48.9° and 94.4°, respectively. These results further indicate that the LJ description of the vdW interactions between water and *h*-BN is not reliable.

*4.2. Pristine Multilayer h-BN*

To investigate the influence of *h*-BN substrate thickness on the WCA, we constructed pristine multilayer AA'-stacked *h*-BN models with varying numbers of layers, ranging from one to five. The interlayer distance between adjacent *h*-BN layers was kept constant at 3.33 Å. During the simulations, we utilized a water droplet containing 4000 water molecules. Following a similar protocol as described in the previous simulations, the droplets underwent an equilibration period of 10 ns (see more details in Sec. 4.1). The WCAs were then calculated based on the data collected during the last 5 ns of the simulation. As depicted in Fig. 5(d), the calculated WCA is almost independent of the number of *h*-BN layers, suggesting a negligible influence of substrate thickness on the WCA.

*4.3. h-BN with Atomic Edges*

Based on the results and discussions in Sec. 4.1 and 4.2, it is evident that there exists a notable discrepancy between the WCAs simulated using our developed AIP (115.5°) and the experimental values (52° − 67°) observed for both monolayer and multilayer *h*-BN.[15, 18, 21] A previous study has attributed this discrepancy to the influence of defects and surface roughness of *h*-BN on its wettability.[19] Their extensive investigations indicate that the presence of atomic step edges on the *h*-BN surface, rather than the presence of defects, plays a substantial role in determining its wettability.

In this section, we aimed to further investigate the effect of surface roughness induced by the atomic step edges on the WCA of *h*-BN. We employed the free-energy perturbation (FEP) method[86, 87] to calculate the WCA, which offers the advantage of not requiring curve fitting and is well-suited for simulating 2D material surfaces with roughness.[19] This method involves determining the work of adhesion of water on *h*-BN ($W_{SL}$) and the surface tension of the water ($\gamma_{lv}$) through MD simulations. Subsequently, the WCA can be estimated using the Young-Dupré equation:

$$W_{SL} = \gamma_{lv}(1 + \cos\theta). \tag{9}$$

In this MD simulation, we constructed a model system with 1000 water molecules deposited on 5 layers of AA'-stacked *h*-BN with a fixed interlayer distance of 3.33 Å, in which an atomic step edge is introduced in the top *h*-BN layer, as shown in Fig. 5(c). Further MD simulation setup and computational details of the FEP method can be found in Section S5 of the SI, from which we obtained $W_{SL} = 95.7$ mJ/m². Following the same protocol presented in Ref. 59, the surface tension of the water on *h*-BN, $\gamma_{lv}$, is calculated as 66.5 mJ/m². Substituting $W_{SL}$ and $\gamma_{lv}$ into Eq. (9), the WCA of the roughened *h*-BN surface with atomic step edges was calculated as $\theta = 64.0°$, which exhibits good agreement with experimental measurements. This finding strongly suggests the presence of surface steps on actual *h*-BN samples is highly likely. Note that we adopted different methods for calculating the WCA of pristine *h*-BN (density profile method) and roughened *h*-BN surface with atomic step edges (FEP method). However, we confirmed that both methods give consistent results for the same model system. For instance, we also applied the FEP method to calculate the WCA for the monolayer pristine *h*-BN and obtained an adhesion work between water and *h*-BN of $W_{SL} = 35.6$ mJ/m², resulting in a WCA of 117.7°. This value is consistent with that obtained from the density profile method (115.5°) (see Sec. 4.1 for details).

Based on the calculation results, we can draw the conclusion that the presence of atomic step edges on the *h*-BN surface significantly enhances the adhesion work between water and *h*-BN (95.7 mJ/m² vs. 35.6 mJ/m²), which in turn leads to a significant decrease in the WCA. Furthermore, we investigated the influence of the number of *h*-BN layers on the WCA for *h*-BN with atomic step edges on the surface. As depicted in Fig. 5(d), we observed that the WCA remains almost unaffected by the substrate thickness. This finding aligns with the conclusion drawn in Section 4.2 for pristine *h*-BN, reinforcing the notion that the substrate thickness has minimal impact on the WCA even when atomic step edges are present.

TABLE 2. Summary of WCA on monolayer and bulk *h*-BN (Samples/Setup), method utilized to measure the WCA in experiments (Method), and the force field parameters employed in simulations to model the interaction of H₂O/*h*-BN heterointerfaces.

| | | WCA | Samples | Method | Refs |
|---|---|---|---|---|---|
| Experiments | | 62.6° | 1L[b] *h*-BN/Cu | Sessile drop | 15 |
| | | 61.4° | 1L *h*-BN/Ge | | |
| | | 65.7° | 1L *h*-BN/Ni | | |
| | | ~80° | 1L *h*-BN/SiO₂/Si | Sessile drop | 17 |
| | | 52° | 1L *h*-BN/SiO₂/Si | Sessile drop | 18 |
| | | 165° | bulk porous *h*-BN | Sessile drop | 1 |
| | | 67° | bulk *h*-BN | Sessile drop | 16 |
| | | 55.3±1.5° | bulk *h*-BN | Sessile drop | 20 |
| | | 40° | bulk *h*-BN | Sessile drop | 88 |
| | | 44°-52° | bulk *h*-BN | Sessile drop | 89 |
| | | 65°-67° | bulk *h*-BN | Sessile drop | 21 |
| MD Simulations | LJ potential | WCA | Setup | Force field parameters[a] | - |
| | | 56° | TIP3P water model/hBN (1L) | $\sigma_{NO}= 3.20, \varepsilon_{NO}= 5.70, \sigma_{NH}= 1.80, \varepsilon_{NH}= 2.69$ $\sigma_{BO}= 3.30, \varepsilon_{BO}= 2.69, \sigma_{BH}= 1.90, \varepsilon_{BH}= 0.62$ | 23 |
| | | 55° | TIP4P water model/bulk *h*-BN | $\sigma_{NM}= 3.28, \varepsilon_{NM}= 5.26, \sigma_{BM}= 3.32, \varepsilon_{BM}= 4.25$ | 25 |

| | | | | |
|---|---|---|---|---|
| | | ~ 0° | SPC-F2 water model/bulk h-BN | $\sigma_{NO}$= 3.27, $\varepsilon_{NO}$= 6.50, $\sigma_{NH}$= 2.79, $\varepsilon_{NH}$= 2.69 $\sigma_{BO}$= 3.31, $\varepsilon_{BO}$= 2.16, $\sigma_{BH}$= 2.84, $\varepsilon_{BH}$= 1.75 | |
| | | 81° | TIP4P/Ice water model/bulk h-BN | $\sigma_{NO}$= 3.19, $\varepsilon_{NO}$= 4.33, $\sigma_{BO}$=3.24, $\varepsilon_{BO}$= 5.24 | 24 |
| | | ~28° | SPC/E water model/h-BN (1L) | $\sigma_{NO}$= 3.27, $\varepsilon_{NO}$= 6.50, $\sigma_{BO}$=3.31, $\varepsilon_{BO}$= 5.26 | 45 |
| | | 15.4±5.1° | TIP4P/Ice water model/bulk h-BN | $\sigma_{NO}$= 3.27, $\varepsilon_{NO}$= 7.58, $\sigma_{BO}$=3.31, $\varepsilon_{BO}$= 6.13 | 44 |
| | | 42.6±0.6° | TIP4P/Ice water model/bulk h-BN | $\sigma_{NO}$= 3.19, $\varepsilon_{NO}$= 8.97, $\sigma_{BO}$=3.27, $\varepsilon_{BO}$= 4.14 | 45 |
| | | 73° | SPC/E water model/h-BN (2L) | $\sigma_{NO}$= 3.27, $\varepsilon_{NO}$= 6.51, $\sigma_{BO}$=3.31, $\varepsilon_{BO}$= 5.27 | 27 |
| | QMD | 86° | h-BN (1L) | - | 46 |
| | This work | 115.5° | TIP4P/2005 water model/pristine h-BN (1L) | See Table S2 of the SI | - |
| | | 114.3° | TIP4P/2005 water model/pristine bulk h-BN (6L) | See Table S2 of the SI | - |
| | | 64.0° | TIP4P/2005 water model/ bulk h-BN with atomic steps | See Table S2 of the SI | - |

[a]For LJ potential, the parameters $\sigma_{N\alpha}$ (Å), $\varepsilon_{N\alpha}$ (meV), $\sigma_{B\alpha}$ (Å), and $\varepsilon_{B\alpha}$ (meV) are provided. Parameters for other potentials are not included. [b]L refers to the number of h-BN layers.

Table 2 provides a comprehensive summary of the WCAs reported for monolayer and multilayer h-BN, obtained through both experiments and simulations. The WCAs were measured (calculated) using various methods (force fields) and different samples (setups). The experimental values of WCAs for monolayer and bulk h-BN are in ranges of 52° − 62.6° and 40° − 67°, respectively. However, the WCAs predicted by the LJ potential range from 0° to 81°, employing a diverse set of parameters (see Table 2). We further show that the LJ potential can only describe one BE curve for a fixed configuration (details are provided in Section S3 of the SI). All the information indicates that the simple isotropic force field cannot fully mimic the anisotropic interactions between water molecules and h-BN. In comparison, the anisotropic interfacial force field developed in this study demonstrates remarkable capability by accurately capturing all DFT reference data using a single parameter set. By employing the developed AIP, we investigated the wettability of atomically-flat pristine h-BN and h-BN with atomic steps, and came to the conclusion that the presence of atomic step edges leads to a transformation of the wettability of h-BN from hydrophobic to hydrophilic due to the enhancement of the adhesion work from the edges. The calculated WCA in the latter case matches well with the experimental value.

## 5. Conclusion

In summary, the presented results provide compelling evidence for the effectiveness of the proposed anisotropic interfacial potential in accurately capturing the energetics of water/*h*-BN heterostructures. Unlike the traditional Lennard-Jones (LJ) potential, the AIP utilizes a single parameter set and successfully reproduces the behavior of the system, as demonstrated by the agreement between the computed various DFT reference BE curves and sliding PESs. MD simulations utilizing the parameterized AIP reveal that the presence of exposed atomic edges enhances the hydrophilic nature of *h*-BN surfaces, while the atomically-flat pristine *h*-BN surface exhibits inherent hydrophobicity. These findings provide valuable insights and may serve as a catalyst for future experimental investigations aimed at determining the intrinsic wettability of atomically-flat pristine *h*-BN. The development of AIP for water/*h*-BN herointerfaces opens up exciting possibilities for accurate and productive simulations of such heterostructures. This advancement provides a valuable tool for conducting comprehensive investigations into *h*-BN's transport, wetting, and tribological properties. Moreover, the methodology proposed here can be readily extended to explore the wettability of other polarized 2D materials, broadening its applicability and facilitating further research in this field.

**SUPPLEMENTARY MATERIAL**

The Supplementary Information accompanying the subsequent sections includes: Convergence Tests of the Reference DFT Calculations, Initial Configurations for Optimization, Assessment of LJ potential describing Van der Waals interaction between water and *h*-BN, Fitting Procedure and AIP Parameters for water/*h*-BN heterostructure, and Calculation of the Work of Adhesion using free-energy perturbation (FEP) method.

**ACKNOWLEDGMENTS**

This work was supported by the Natural Science Foundation of Hubei Province (No. 2021CFB138) and the National Natural Science Foundation of China (No. 12102307), the Key Research and Development Program of Hubei Province (No. 2021BAA192), and the Fundamental Research Funds for the Central Universities (No. 2042022kf1177), and the start-up fund of Wuhan University.

# References

[1] W. Lei, D. Portehault, D. Liu, S. Qin, and Y. Chen, Nat. Commun. **4**, 1777 (2013)

[2] J. Azamat, A. Khataee, and S. W. Joo, J. Mol. Model. **22**, 82 (2016)

[3] B. B. Sharma, and A. Govind Rajan, J. Phys. Chem. B **126**, 1284 (2022)

[4] H. Gao, Q. Shi, D. Rao, Y. Zhang, J. Su, Y. Liu, Y. Wang, K. Deng, and R. Lu, J. Phys. Chem. C **121**, 22105 (2017)

[5] X. Davoy, A. Gellé, J.-C. Lebreton, H. Tabuteau, A. Soldera, A. Szymczyk, and A. Ghoufi, ACS Omega **3**, 6305 (2018)

[6] K. Zhang, Y. Feng, F. Wang, Z. Yang, and J. Wang, Journal of Materials Chemistry C **5**, 11992 (2017)

[7] M. B. Panchal, and S. H. Upadhyay, IET Nanobiotechnol. **8**, 149 (2014)

[8] S. Liu, B. Lu, Q. Zhao, J. Li, T. Gao, Y. Chen, Y. Zhang, Z. Liu, Z. Fan, F. Yang, L. You, and D. Yu, Adv. Mater. **25**, 4549 (2013)

[9] X. Li, S. Chen, Q. Liu, Y. Luo, and X. Sun, Chem. Commun. **57**, 8039 (2021)

[10] K.-H. Oh, D. Lee, M.-J. Choo, K. H. Park, S. Jeon, S. H. Hong, J.-K. Park, and J. W. Choi, ACS Appl. Mater. Interfaces **6**, 7751 (2014)

[11] D. Chimene, D. L. Alge, and A. K. Gaharwar, Adv. Mater. **27**, 7261 (2015)

[12] A. Siria, P. Poncharal, A.-L. Biance, R. Fulcrand, X. Blase, S. T. Purcell, and L. Bocquet, Nature **494**, 455 (2013)

[13] A. Pendse, S. Cetindag, P. Rehak, S. Behura, H. Gao, N. H. L. Nguyen, T. Wang, V. Berry, P. Král, J. Shan, and S. Kim, Adv. Funct. Mater. **31**, 2009586 (2021)

[14] A. Pakdel, C. Zhi, Y. Bando, T. Nakayama, and D. J. A. N. Golberg, **5**, 6507 (2011)

[15] X. Li, H. Qiu, X. Liu, J. Yin, and W. Guo, Adv. Funct. Mater. **27**, 1603181 (2017)

[16] G.-X. Li, Y. Liu, B. Wang, X.-M. Song, E. Li, and H. Yan, Appl. Surf. Sci. **254**, 5299 (2008)

[17] A. Keerthi, S. Goutham, Y. You, P. Iamprasertkun, R. A. W. Dryfe, A. K. Geim, and B. Radha, Nat. Commun. **12**, 3092 (2021)

[18] W. Sheng, I. Amin, C. Neumann, R. Dong, T. Zhang, E. Wegener, W. L. Chen, P. Forster, H. Q. Tran, M. Loffler, A. Winter, R. D. Rodriguez, E. Zschech, C. K. Ober, X. Feng, A. Turchanin, and R. Jordan, Small **15**, e1805228 (2019)

[19] A. Kumar Verma, and A. Govind Rajan, Langmuir **38**, 9210 (2022)

[20] L. B. Boinovich, A. M. Emelyanenko, A. S. Pashinin, C. H. Lee, J. Drelich, and Y. K. Yap, Langmuir **28**, 1206 (2012)

[21] V. A. S. Kandadai, V. Gadhamshetty, and B. K. Jasthi, Surf. Coat. Technol. **447**, 128805 (2022)

[22] A. Verma, W. Zhang, and A. C. T. van Duin, Phys. Chem. Chem. Phys. **23**, 10822 (2021)

[23] A. Budi, and T. R. Walsh, Langmuir **35**, 16234 (2019)

[24] A. Govind Rajan, M. S. Strano, and D. Blankschtein, J. Phys. Chem. Lett. **9**, 1584 (2018)

[25] Y. Wu, L. K. Wagner, and N. R. Aluru, J. Chem. Phys. **144**, 164118 (2016)

[26] L. Garnier, A. Szymczyk, P. Malfreyt, and A. Ghoufi, J. Phys. Chem. Lett. **7**, 3371 (2016)

[27] R. C. Dutta, S. Khan, and J. K. Singh, Fluid Phase Equilib. **302**, 310 (2011)

[28] S. Y. Misyura, V. A. Andryushchenko, D. V. Smovzh, and V. S. Morozov, Materials Science and Engineering: B **277**, 115588 (2022)

[29] Y. Lv, P. L. Chong, and S. Liu, Materials Science in Semiconductor Processing **142**, 106452 (2022)

[30] A. Tuoliken, L. Zhou, P. Bai, and X. Du, Int. J. Heat Mass Transfer **172**, 121218 (2021)

[31] J. Włoch, A. P. Terzyk, and P. Kowalczyk, Chem. Phys. Lett. **674**, 98 (2017)

[32] M. Kanduc, J. Chem. Phys. **147**, 174701 (2017)

[33] C. Huang, F. Xu, and Y. Sun, Comput. Mater. Sci. **139**, 216 (2017)

[34] J. Wloch, A. P. Terzyk, P. A. Gauden, R. Wesolowski, and P. Kowalczyk, J. Phys.: Condens. Matter **28**, 495002 (2016)

[35] J. E. Andrews, S. Sinha, P. W. Chung, and S. Das, Phys. Chem. Chem. Phys. **18**, 23482 (2016)

[36] F. Leroy, S. Liu, and J. Zhang, J. Phys. Chem. C **119**, 28470 (2015)

[37] S. Becker, H. M. Urbassek, M. Horsch, and H. Hasse, Langmuir **30**, 13606 (2014)

[38] Y. Wu, and N. R. Aluru, J. Phys. Chem. B **117**, 8802 (2013)

[39] E. Santiso, C. Herdes, and E. Müller, Entropy **15**, 3734 (2013)

[40] G. Perez-Hernandez, and B. Schmidt, Phys. Chem. Chem. Phys. **15**, 4995 (2013)

[41] D. Sergi, G. Scocchi, and A. Ortona, Fluid Phase Equilib. **332**, 173 (2012)

[42] H. Peng, A. V. Nguyen, and G. R. Birkett, Molecular Simulation **38**, 945 (2012)

[43] A. Seal, and A. Govind Rajan, Nano Lett. **21**, 8008 (2021)

[44] S. L. Mayo, B. D. Olafson, and W. A. J. J. o. P. c. Goddard, **94**, 8897 (1990)

[45] C. Y. Won, and N. R. Aluru, J. Am. Chem. Soc. **129**, 2748 (2007)

[46] H. Li, and X. C. Zeng, ACS Nano **6**, 2401 (2012)

[47] J. Zhang, K. Jia, Y. Huang, X. Liu, Q. Xu, W. Wang, R. Zhang, B. Liu, L. Zheng, H. Chen, P. Gao, S. Meng, L. Lin, H. Peng, and Z. Liu, Adv. Mater. **34**, e2103620 (2022)

[48] J. Ma, A. Michaelides, D. Alfe, L. Schimka, G. Kresse, and E. Wang, Phys. Rev. B **84**, 033402 (2011)

[49] E. R. Cortes, L. M. Solís, and J. Arellano, Rev. Mex. Fis. **59**, 118 (2013)

[50] Y. S. Al-Hamdani, M. Ma, D. Alfe, O. A. von Lilienfeld, and A. Michaelides, J. Chem. Phys. **142**, 181101 (2015)

[51] Y. S. Al-Hamdani, M. Rossi, D. Alfe, T. Tsatsoulis, B. Ramberger, J. G. Brandenburg, A. Zen, G. Kresse, A. Gruneis, A. Tkatchenko, and A. Michaelides, J. Chem. Phys. **147**, 044710 (2017)

[52] K. Huang, H. Qin, S. Zhang, Q. Li, W. Ouyang, and Y. Liu, Adv. Funct. Mater. **32**, 2204209 (2022)

[53] W. Ouyang, H. Qin, M. Urbakh, and O. Hod, Nano Lett. **20**, 7513 (2020)

[54] W. Ouyang, O. Hod, and M. Urbakh, Phys. Rev. Lett. **126**, 216101 (2021)

[55] W. Ouyang, D. Mandelli, M. Urbakh, and O. Hod, Nano Lett. **18**, 6009 (2018)

[56] W. Ouyang, R. Sofer, X. Gao, J. Hermann, A. Tkatchenko, L. Kronik, M. Urbakh, and O. Hod, J. Chem. Theory Comput. **17**, 7237 (2021)

[57] W. Ouyang, O. Hod, and R. Guerra, J. Chem. Theory Comput. **17**, 7215 (2021)

[58] W. Ouyang, I. Azuri, D. Mandelli, A. Tkatchenko, L. Kronik, M. Urbakh, and O. Hod, J. Chem. Theory Comput. **16**, 666 (2020)

[59] Z. Feng, Y. Yao, J. Liu, B. Wu, Z. Liu, and W. Ouyang, J. Phys. Chem. C **127**, 8704 (2023)

[60] P. E. Blöchl, Phys. Rev. B **50**, 17953 (1994)

[61] J. Sun, A. Ruzsinszky, and J. P. Perdew, Phys. Rev. Lett. **115**, 036402 (2015)

[62] T. Maaravi, I. Leven, I. Azuri, L. Kronik, and O. Hod, J. Phys. Chem. C **121**, 22826 (2017)

[63] I. Leven, T. Maaravi, I. Azuri, L. Kronik, and O. Hod, J. Chem. Theory Comput. **12**, 2896 (2016)

[64] I. Leven, I. Azuri, L. Kronik, and O. Hod, J. Chem. Phys. **140**, 104106 (2014)

[65] A. N. Kolmogorov, and V. H. Crespi, Phys. Rev. B **71**, 235415 (2005)

[66] E. de Vos Burchart, V. A. Verheij, H. van Bekkum, and B. van de Graaf, Zeolites **12**, 183 (1992)

[67] T. Tian, and C.-J. Shih, Ind. Eng. Chem. Res. **56**, 10552 (2017)

[68] A. N. Kolmogorov, and V. H. J. P. R. L. Crespi, Phys. Rev. Lett. **85**, 4727 (2000)

[69] A. Tkatchenko, and M. Scheffler, Phys. Rev. Lett. **102**, 073005 (2009)

[70] J. L. Abascal, and C. Vega, J. Chem. Phys. **123**, 234505 (2005)

[71] R. H. Byrd, J. C. Gilbert, and J. Nocedal, Math. Program. **89**, 149 (2000)

[72] R. A. Waltz, J. L. Morales, J. Nocedal, and D. Orban, Math. Program. **107**, 391 (2006)

[73] S. Plimpton, J. Comput. Phys. **117**, 1 (1995)

[74] S. Nosé, J. Chem. Phys. **81**, 511 (1984)

[75] A. Stukowski, Modell. Simul. Mater. Sci. Eng. **18**, 015012 (2009)

[76] P. Dančová, V. Vinš, D. Celný, B. Planková, T. Němec, M. Duška, J. Hrubý, and M. Veselý, EPJ Web Conf. **114**, 02136 (2016)

[77] S. P. Kadaoluwa Pathirannahalage, N. Meftahi, A. Elbourne, A. C. G. Weiss, C. F. McConville, A. Padua, D. A. Winkler, M. Costa Gomes, T. L. Greaves, T. C. Le, Q. A. Besford, and A. J. Christofferson, J. Chem. Inf. Model. **61**, 4521 (2021)

[78] C. Vega, and E. de Miguel, J. Chem. Phys. **126**, 154707 (2007)

[79] K. Ritos, N. Dongari, M. K. Borg, Y. Zhang, and J. M. Reese, Langmuir **29**, 6936 (2013)

[80] C. Vega, and J. L. Abascal, Phys. Chem. Chem. Phys. **13**, 19663 (2011)

[81] I. Essafri, J.-C. Le breton, A. Saint-Jalmes, A. Soldera, A. Szymczyk, P. Malfreyt, and A. Ghoufi, Molecular Simulation **45**, 454 (2018)

[82] J.-P. Ryckaert, G. Ciccotti, and H. J. C. Berendsen, J. Comput. Phys. **23**, 327 (1977)

[83] R. W. Hockney, and J. W. Eastwood, *Computer simulation using particles* (crc Press, 2021)

[84] M. Ma, G. Tocci, A. Michaelides, and G. Aeppli, Nat. Mater. **15**, 66 (2016)

[85] X. Zhang, F. Shi, J. Niu, Y. Jiang, and Z. Wang, J. Mater. Chem. **18**, 621 (2008)

[86] M. R. Shirts, and J. D. Chodera, J. Chem. Phys. **129**, 124105 (2008)

[87] R. W. Zwanzig, J. Chem. Phys. **22**, 1420 (1954)

[88] J. Yu, L. Qin, Y. Hao, S. Kuang, X. Bai, Y.-M. Chong, W. Zhang, and E. Wang, ACS Nano **4**, 414 (2010)

[89] C. Lee, J. Drelich, and Y. Yap, Langmuir **25**, 4853 (2009)